# Wide-band X-ray variability of GRS 1915+105 observed with BeppoSAX

P. Casella*, M. Feroci*, E. Massaro*, E. Costa*, M. Litterio*, G.Matt[†], T. Belloni**, M. Tavani[‡], I.F. Mirabel[§], A.J. Castro-Tirado[¶], A. Harmon[∥] and G. Pooley[††]

*Istituto di Astrofisica Spaziale. CNR, Roma, Italy
[†]Physics Department, Universitá Roma TRE, Roma, Italy
**Osservatorio Astronomico di Brera, Merate, Italy
[‡]Istituto di Fisica Cosmica, CNR, Milano, Italy
[§]Service d' Astrophysique, CEA, Saclay, France
[¶]Istituto de Astrofisica de Andalusia, Granada, Spain
[∥]NASA Marshall Space Flight Center, Hunstville, Alabama, USA
[††]Mullard Radio Astronomy Observatory, Cambridge, UK

**Abstract.**
   The Galactic Microquasar GRS 1915+105 was observed by the Narrow Field Instruments onboard BeppoSAX in two pointings during year 2000. We present the preliminary results of wide-band study of the observed variability, carried out also with the wavelet analysis.

## INTRODUCTION

Since its discovery in 1992 (Castro-Tirado, Brandt & Lund 1992), GRS 1915+105, the prototype of galactic microquasars (Mirabel & Rodriguez 1994), is one of the most observed sources, particularly in the X-ray range. Its light curve shows a high variable behavior, characterized by strong and structured outbursts interrupted by quiescent phases. No evidence of periodic variations, related to orbital motion in a binary system, has been firmly established. It is general opinion, by the analogy with similar sources, that the X-ray emission is originated in an accretion disk rotating around a few stellar mass black hole. The X-ray spectrum is generally fitted by at least two components: a multi-temperature disk blackbody and a power law extending to several hundreds keV. During the flares the main spectral parameters of the thermal component show significant variations, which have been interpreted by the emptying and refilling of the inner portion of the accretion disk (Belloni et al. 1997).

In a recent paper, Belloni et al. (2000), on the basis of a large set of RXTE observations, defined 12 different variability modes of the X-ray emission, each of them characterized by a time profile and spectral variability as apparent from the dynamical hardness ratio plots. This classification is potentially useful for the understanding of the physical processes occurring in this exceptional source and will be used in our analysis.

In this contribution we present some preliminary results of a wide band X-ray observations of GRS 1915+105 recently performed with the Italian-Dutch satellite BeppoSAX.

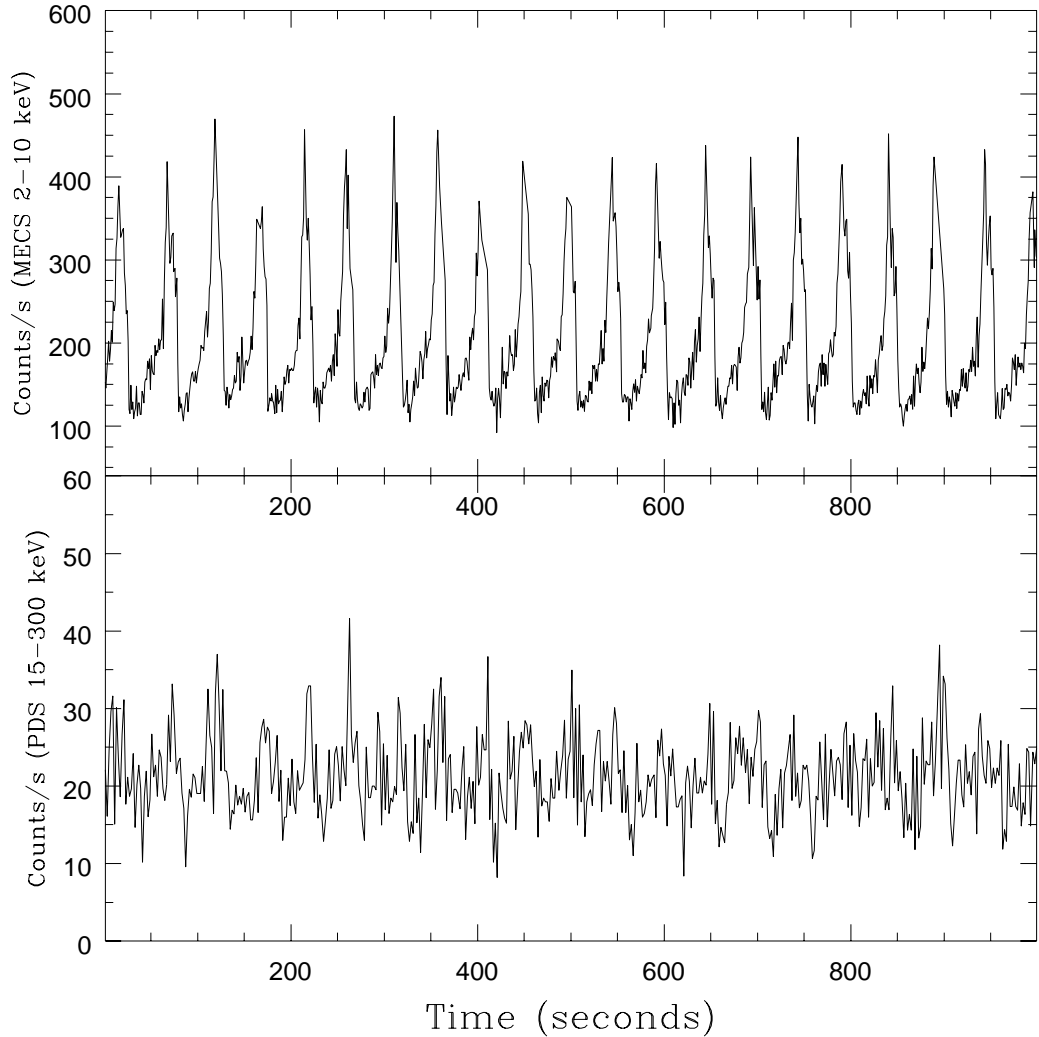

**FIGURE 1.** Two simultaneous portions of the light curve in the MECS and PDS during a regular ρ mode.

In particular, we compare the light curves in the energy bands of the MECS (2-10 keV) and PDS (13-300 keV) in regular and irregular variabiltiy modes and describe the evolution of short time scale by means of wavelet power spectra.

## OBSERVATIONS AND DATA REDUCTION

Recent BeppoSAX observational campaigns on GRS 1915+105 were performed in the period April 1999, April and October 2000. The main parameters of the last three observations are reported in Table 1. Standard procedures and selection criteria were applied to the BeppoSAX data to avoid the South Atlantic Anomaly, solar, bright Earth and particle contamination using the SAXDAS v. 2.0.0 package. Source counts for each energy channel were extracted from the LECS and MECS images using standard algorithms and parameters. PDS data were taken in the standard direct mode, with a collimator rocking law with 96 s dwell time in order to observe simultaneously source and background.

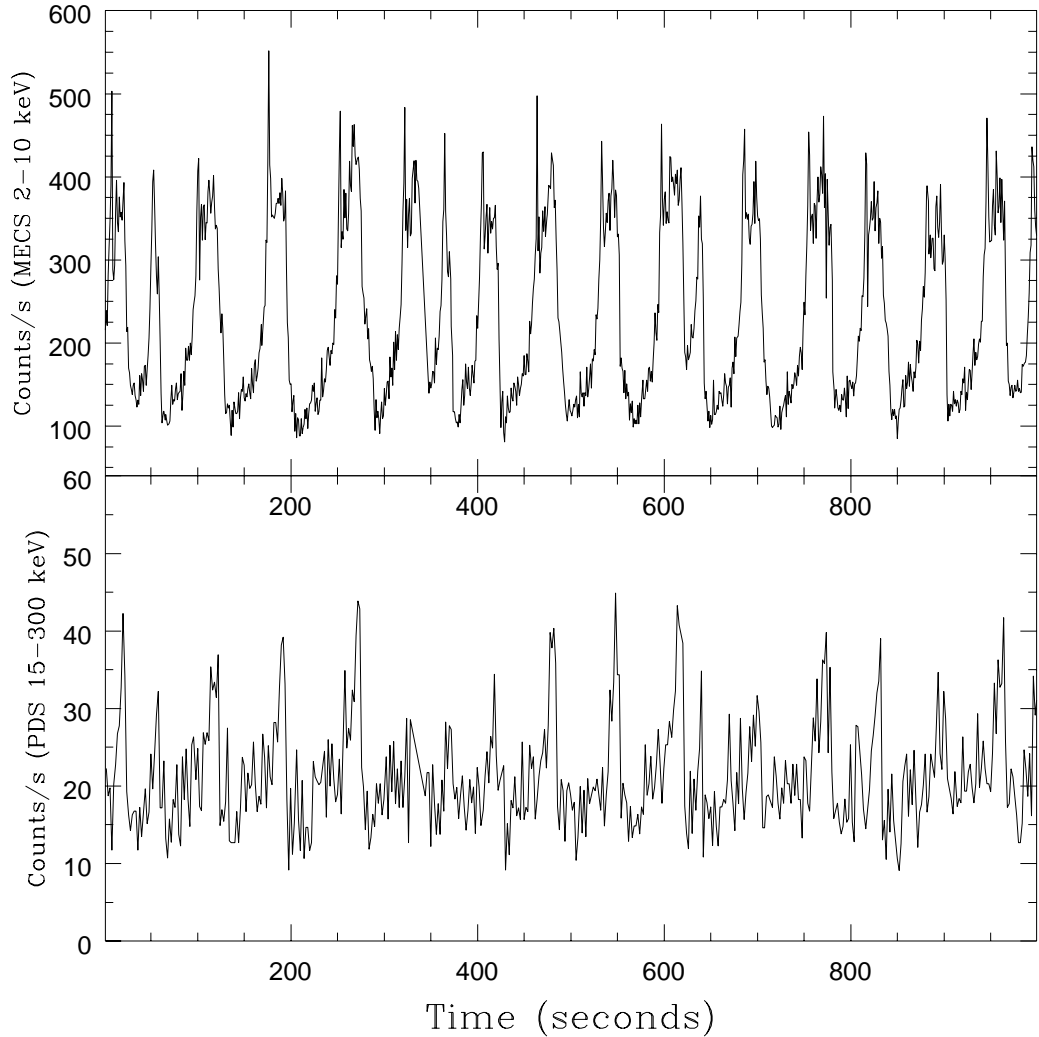

**FIGURE 2.** Two simultaneous portions of the light curve in the MECS and PDS during a non-regular ρ mode.

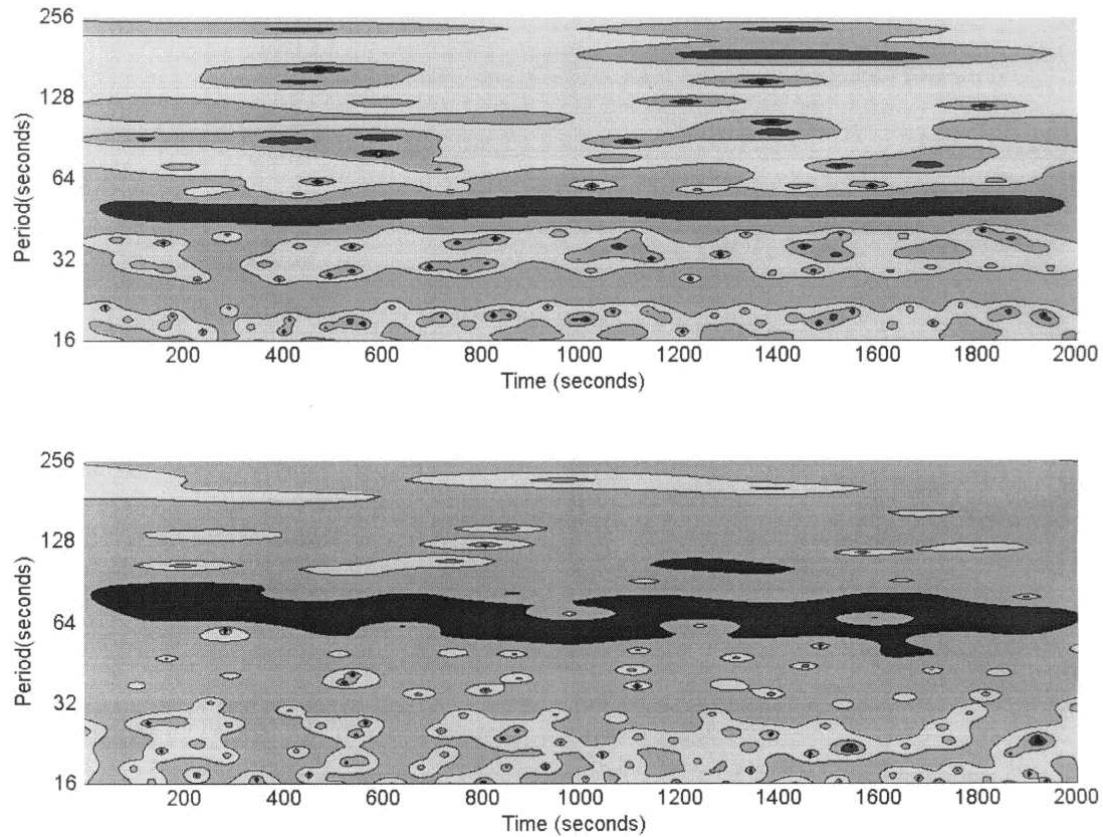

**FIGURE 3.** Wavelet power spectra of two MECS light curve portions (each 2,000 s long) during regular (Fig. 1 time series, upper panel) and non-regular (Fig. 2 time series, lower panel) ρ mode.

# THE STRUCTURE OF THE LIGHT CURVES

For the entire duration of the April 1999 observation the 2-10 keV X-ray emission of GRS 1915+105 showed the characteristic pulsed modulation of the ρ mode (Belloni et al. 2000). Pulses were not well evident in the PDS (13-300 keV) and LECS (0.1-2 keV) bands because of the lower count rate due to the strong interstellar absorption below a few keV and the dominance of the the power law component above 15 keV (for a detailed analysis see e.g. Casella et al. 2001, Feroci et al. 2001).

The April 2000 pointing covered only about 3 days in the course of a longer (10 days) multifrequency campaign (Ueda et al. 2001). In the MECS and PDS time series the X-ray emission of GRS 1915+105 was characterized by a slow variability superimposed to a random noise, whose amplitude was compatible with the Poisson statistical fluctuations. A large flare was detected in the MECS, but not in the PDS. Unfortunately, it was just at the end of the visibility window during the BeppoSAX orbit and therefore its time evolution was not entirely followed.

The observation of October 2000 was particularly important because of its long overall duration, about 10 days, which gave us the possibility to detect the transition between different modes. A stable ρ mode was observed during the first 160,000 s followed by an irregular phase about 25,000 s long. Another and longer not stable phase was in the interval from about 275,000 s to 375,000 s and after the source showed again for a long time a stable ρ mode interrupted by a quiescent state at 615,000 s. Finally, in the final part, a transition to the ν mode occurred. Two pairs of short segments, 1,000 s long, of the MECS and PDS light curves are shown in the Fig. 1 and 2, respectively. The former, corresponds to the regular ρ mode, while the latter to the irregular one, which is characterized by broader pulses with a recurrence time longer than 70 s, and by the apparence of narrower pulses between two main ones. The higher energy time series are also quite different: pulses are barely detectable in Fig. 1, while they are well evident in Fig. 2.

The stability of these two types of behavior of the ρ mode was also studied by means of the wavelet analysis. Wavelet power spectra were computed using the Morlet wavelet of order 12 (Farge 1992). Two spectra for the MECS time series are presented in the panels of Fig. 3. The spectrum of the regular ρ mode is characterized by a rather stable recurrence time of the peaks of about 50 s, as shown by an uninterrupted dark linear strip (corresponding to the highest power), centered at this period value. In the irregular mode this strip shows meanderings and in some cases more dominant frequencies appear at the same time.

An important finding is that the pulses in the high energy curve occurr in the second half of the corresponding low energy peaks. This effect suggest that there is a delay like the emission would be due to different components. Such behavior is reminiscent of that observed in the so called *plateau* intervals, during which a QPO frequency lower than 2 Hz is observed (Reig et al. 2000), and interpreted by Nobili et al. (2000) in terms of a comptonization model. Another possibility is that the non-regular ρ mode is associated with a greater oscillation amplitude of the innermost region of the accretion disk where higher temperatures can be reached. A more complete time resolved spectral analysis, however, is necessary to derive further information on the physical conditions of the emitting plasma and its instabilities.